\documentclass[utf8,final]{FrontiersinVancouver}

\usepackage{url,hyperref,microtype,subcaption}
\usepackage[onehalfspacing]{setspace}
\usepackage{tcolorbox}
\usepackage{booktabs,tabularx}
\usepackage{array}
\usepackage{float}

\usepackage{ragged2e}
\newcolumntype{L}[1]{>{\RaggedRight\arraybackslash}p{#1}}
\usepackage{makecell}

\newcommand{\Oplus}{\ensuremath{\vcenter{\hbox{\scalebox{1.7}{$\oplus$}}}}}

\def\keyFont{\fontsize{8}{11}\helveticabold }
\def\firstAuthorLast{Malpetti {et~al.}}
\def\Authors{
  Daniele Malpetti\,$^{1,2,\dagger}$,
  Marco Scutari\,$^{1,\dagger,*}$
  Francesco Gualdi\,$^{1,2,\dagger}$
  Jessica van Setten\,$^{3}$
  Sander van der Laan\,$^{4,5}$
  Saskia Haitjema\,$^{4}$
  Aaron Mark Lee\,$^{6}$
  Isabelle Hering\,$^{7}$
  and Francesca Mangili\,$^{1,2}$
}

\def\Address{
  $^{1}$Istituto Dalle Molle di Studi sull'Intelligenza Artificiale (IDSIA),
    USI-SUPSI, Lugano, Switzerland \\
  $^{2}$Swiss Institute of Bioinformatics (SIB), Lugano, Switzerland \\
  $^{3}$Department of Cardiology, University Medical Center Utrecht,
    University of Utrecht, Utrecht, The Netherlands \\
  $^{4}$Central Diagnostics Laboratory, University Medical Center Utrecht,
    University of Utrecht, Utrecht, The Netherlands \\
  $^{5}$Department of Genome Sciences, University of Virginia, Charlottesville,
    VA, United States \\
  $^{6}$William Harvey Research Institute, NIHR Barts Biomedical Research
    Centre, Queen Mary University of London, London, United Kingdom \\
  $^{7}$Étude Hering, DPO Associates SARL, Nyon, Switzerland \\
  $^{\dagger}$ Equal contributions.
}

\def\corrAuthor{Marco Scutari, Istituto Dalle Molle di Studi sull'Intelligenza
  Artificiale (IDSIA), USI-SUPSI, Polo Universitario Lugano, Via La Santa 1,
  Lugano, 6962, Switzerland}

\def\corrEmail{scutari@bnlearn.com}

\begin{document}
\onecolumn
\firstpage{1}

\title[Federated learning in bioinformatics]
  {Technical and legal aspects of federated learning in bioinformatics:
   applications, challenges and opportunities}

\author[\firstAuthorLast ]{\Authors}
\address{}
\correspondance{}

\extraAuth{}

\maketitle

\begin{abstract}
  Federated learning leverages data across institutions to improve clinical
  discovery while complying with data-sharing restrictions and protecting
  patient privacy. This paper provides a gentle introduction to this approach in
  bioinformatics, and is the first to review key applications in proteomics,
  genome-wide association studies (GWAS), single-cell and multi-omics studies in
  their legal as well as methodological and infrastructural challenges. As the
  evolution of biobanks in genetics and systems biology has proved, accessing
  more extensive and varied data pools leads to a faster and more robust
  exploration and translation of results. More widespread use of federated
  learning may have a similar impact in bioinformatics, allowing academic and
  clinical institutions to access many combinations of genotypic, phenotypic and
  environmental information that are undercovered or not included in existing
  biobanks.

  \tiny
  \keyFont{ \section{Keywords:} Federated machine learning, Exposome, Secure
    distributed analysis, Data privacy, Collaborative genomics}
\end{abstract}

\section{Introduction}
\label{sec:introduction}

Sharing personal information has been increasingly regulated in both the EU
\citep[with the GDPR and the AI act;][]{gdpr,eu-ai} and the US \citep[with HIPAA
and the National AI Initiative Act;][]{hipaa,ai-act} to mitigate the personal
and societal risks associated with their use, particularly in connection with
machine learning and AI models \citep{cath}. These regulations make multi-centre
studies and similar endeavours more challenging, impacting biomedical and
clinical research.

Federated learning \citep[FL;][]{fedavg,ludwig} is a technical solution intended
to reduce the impact of these restrictions. FL allows multiple parties to
collaboratively train a global machine learning model using their respective
data without sharing it themselves, and without any meaningful model performance
degradation. Instead, parties only share model updates, making it impractical to
reconstruct personal information when the appropriate secure computational
measures are implemented \citep{wainakh}.

This approach strengthens \emph{security} by keeping sensitive information
local, improves \emph{privacy} by minimising data exposure even between the
parties involved, and limits \emph{risk} of data misuse by allowing each party
to retain complete control over its data \citep{truong}. If enough parties are
involved, FL may access larger and more varied data pools than centralised
biobanks can provide. This is particularly true if there are legal (or other)
barriers to data centralisation, resulting in more accurate and robust models
than those produced by any individual party.

\begin{figure}[p]
  \centering
  \includegraphics[width=0.92\linewidth]{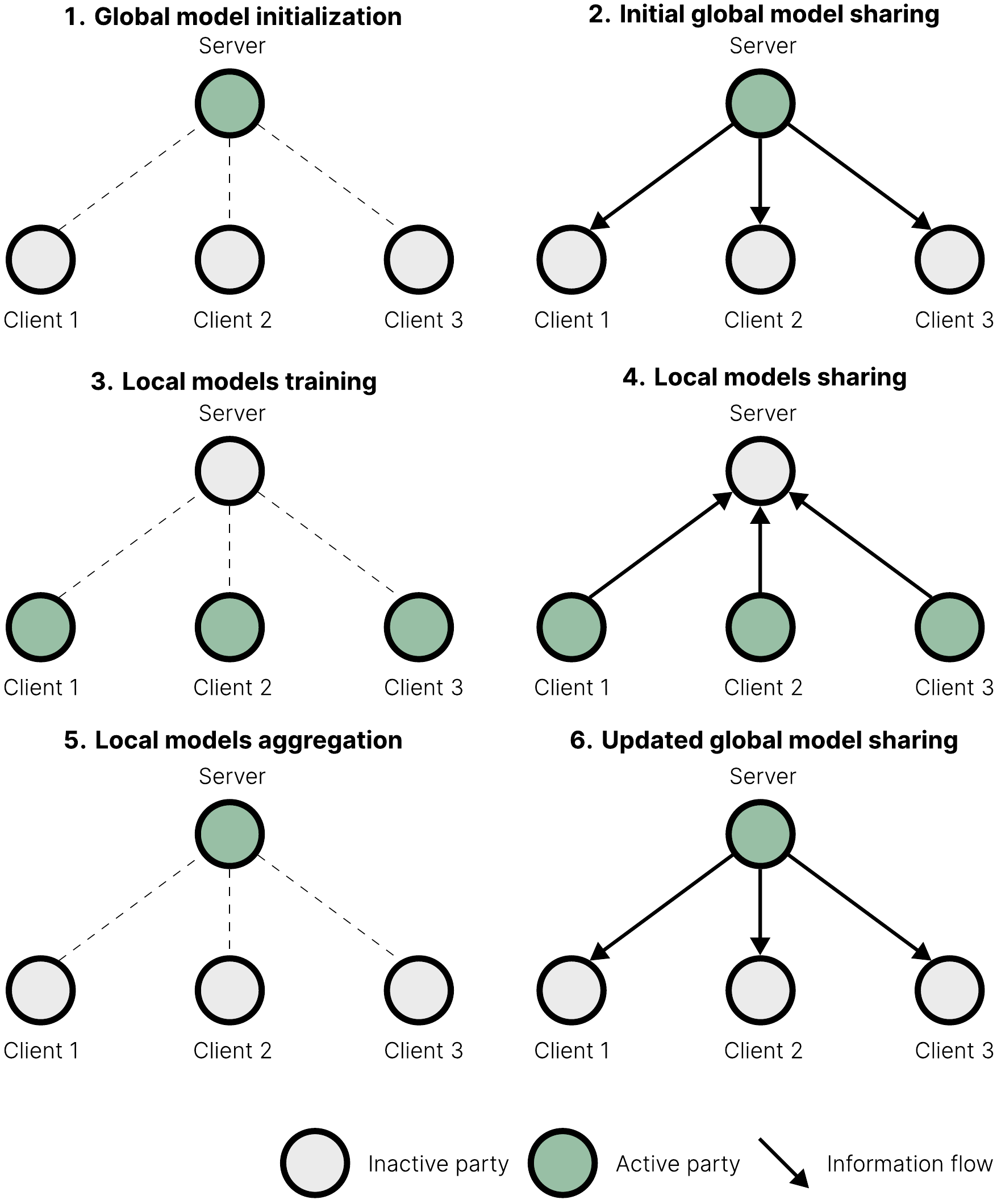}
  \caption{Overview of a typical federated learning (FL) workflow. (1) The
    central server initialises a \emph{global model}. (2) The server shares the
    \emph{global model} parameters with consortium parties, referred to as
    clients. (3) Each client initialises a \emph{local model} from the global
    model parameters and updates it by training it on its local data. (4)
    Clients send their updated \emph{local model} parameters back to the server.
    (5) The server aggregates local model parameters it collected to construct a
    new \emph{global model}. (6) The server redistributes the updated
    \emph{global model} parameters to clients to start the next training round.
    Steps (3)--(6) are repeated iteratively until a predefined stopping
    criterion is met. Active parties in each step are in green, and the arrows
    show the direction of information flow within the consortium.}

\label{fig:intro}
\end{figure}

FL has proven to be a valuable tool for biomedical research and is expected to
gain further traction in the years to come. Its use has improved breast density
classification models \citep[accuracy up by 6\%, generalisability up by
46\%;][]{breast}, COVID-19 outcome prediction at both 24h and 72h \citep[up 16\%
and 38\%;][]{covid19} and rare tumour segmentation \citep[up by 23-33\% and
15\%;][]{pati22} compared to single-party analyses. A consortium of ten
pharmaceutical companies found that FL improved structure-activity relationship
(QSAR) models for drug discovery \citep[both up 12\%][]{melloddy}. Early-stage
applications building predictive models from electronic health records
\citep{brisimi18} have also confirmed no practical performance degradation
compared to pooling data from all parties.

To achieve such results, a real-world implementation of FL must overcome several
methodological, infrastructural and legal issues. However, biomedical FL
literature reviews \citep[among others]{xu20,chowdhury22} are predominantly
high-level and considered simulated rather than real-world implementations.
Here, we will cover federated methods designed explicitly for bioinformatics and
discuss the infrastructure they require, as well as how they meet legal
requirements. In reviewing the literature, we selected papers that study
practical analysis problems (as opposed to proposing methodologies in the
abstract) for proteomics, genome-wide association studies (GWAS), and
single-cell and multi-omics data. We also considered papers that discuss their
feasibility, trade-offs, and performance compared to centralised analyses, and
were published after 2016. We used Google Scholar to find and retrieve them.

To this end, we have structured the remainder of the paper as follows: We first
review the fundamental concepts and design decisions of FL in
Section~\ref{sec:foundations}, including different topologies
(Section~\ref{sec:topologies}), hardware and software (Section~\ref{sec:infra}),
data layouts in different parties (Sections~\ref{sec:allocations}
and~\ref{sec:partitioning}), security (Section~\ref{sec:security}) and privacy
concerns (Section~\ref{sec:privacy}). In Section~\ref{sec:bioinfofl}, we
contrast and compare bioinformatics FL methods for proteomics and differential
expression (Section~\ref{sec:diffexp}), genome-wide association studies (GWAS;
Section~\ref{sec:gwas}), single-cell RNA sequencing (Section~\ref{sec:rnaseq}),
multiomics (Section~\ref{sec:omics}) and medical imaging
(Section~\ref{sec:imaging}) applications. We conclude the section with notable
examples of ready-to-use software tools (Section~\ref{sec:specialised}).
Section~\ref{sec:operations} provides examples of federated operations common in
bioinformatics. Finally, we discuss the legal implications of using FL
(Section~\ref{sec:legal}) before summarising our perspective in
Section~\ref{sec:conclusions}.

\section{Federated learning}
\label{sec:foundations}

FL is a collaborative approach to machine learning model training, where
multiple institutions form a consortium to jointly train a shared model by
exchanging model updates rather than individual patient data. Typically, FL
involves data holders (called "clients") sharing their local contributions with
a server \citep{fedavg} as outlined in Figure~\ref{fig:intro}. The server then
creates and shares back a global model, inviting the data holders to update and
resubmit their contributions. This process is iterative and involves several
rounds of model update exchanges. Unlike traditional centralised computing, FL
does not store patient data in a central location. Instead, patient data remain
under the control of the respective data owners at their sites, enhancing
privacy.

FL has similarities with \emph{distributed computing}, \emph{meta-analysis}, and
\emph{trusted research environments} (TREs), but also has key differences, which
we highlight below. Table~\ref{tab:centralized_fl_dc_meta_tre} provides a
comparative overview of these approaches.

Distributed computing (DC) \citep{zomaya} divides a computational task among
multiple machines to enhance processing speed and efficiency. Typically, DC
starts from a centrally managed data set spread across multiple machines, which
is assumed to contain independent and identically distributed observations. Each
machine is tasked to process a comparable quantity of data. In contrast, clients
independently join FL with their locally held data, which may vary significantly
in quantity and distribution. While sharing some techniques with FL, distributed
computing aims for computational efficiency and lacks its privacy focus.

On the other hand, meta-analyses \citep{toro} aggregate results across
previously completed studies using statistical methods to account for their
variations, thus allowing researchers to synthesise findings without accessing
personal data and preserve the privacy of individual data sets. Here, FL
collaboratively trains a joint model using distributed data to iteratively
update it while meta-analysis constructs it in a single step from the
pre-existing results. Multiple studies on sequencing data have demonstrated that
FL produces results closer to centralised analysis than from meta-analysis
\citep{sfkit,flimma}.

TREs \citep{tres} provide access to data within a controlled, secure computing
environment for conducting analyses, almost always disallowing data sharing.
Some TREs have a centralised data location and governance; an example is the
Research Analysis Platform (RAP), the TRE for the UK Biobank
\citep[UKB;][]{UKB}. Others, such as FEGA \citep{FEGA}, are decentralised. Each
institution maintains its data locally; only the relevant data are securely
transferred to the computing environment when the analysis is authorised. Unlike
FL, the learning process is not distributed across the data holders. Thus, the
trade-off between TREs and FL is between a centralised, trusted entity with
extensive computational facilities that can place substantial restrictions on
the analysis, and a consortium that requires all parties to apply governance
guidelines and provide compute, but can scale both data access and privacy
guarantees.

\begin{table}[H]
\centering
\caption{Methodological Comparison of Centralised Learning, Federated
  Learning, Distributed Computing, Meta-analysis, and Trusted Research
  Environments. Consent from data subjects is assumed for data use.}
\vspace{0.5\baselineskip}
\label{tab:centralized_fl_dc_meta_tre}
\footnotesize
\resizebox{0.95\textwidth}{!}{
\begin{tabular}{
  >{\raggedright\arraybackslash}p{1.8cm}
  >{\raggedright\arraybackslash}p{2.8cm}
  >{\raggedright\arraybackslash}p{2.8cm}
  >{\raggedright\arraybackslash}p{2.8cm}
  >{\raggedright\arraybackslash}p{2.8cm}
  >{\raggedright\arraybackslash}p{2.8cm}}
\toprule
\textbf{Aspect} & \textbf{Centralised Learning} & \textbf{Federated Learning} &
\textbf{Distributed Computing} & \textbf{Meta-analysis} & \textbf{Trusted Research Environments (TRE)} \\
\midrule
\textbf{Primary goal} &
Aggregate all individual data into one place and train or analyse centrally. &
Collaborative model training across parties without sharing individual data. &
Increase speed and scalability; job parallelisation. &
Combine evidence from completed studies. &
Provide secure, auditable access to sensitive data for research. \\
\midrule
\textbf{Where individual data live} &
Single central repository. &
Stay local at each device/institution. &
Centrally stored, sharded across nodes. &
Remain with original studies; not pooled. &
In a secure environment or under local (federated) control (only relevant data are transferred). \\
\midrule
\textbf{How learning happens} &
Training/analysis is run on pooled data in one environment. &
Participants compute local model updates and send them for secure aggregation in iterative rounds. &
Tasks are partitioned and executed in parallel; results are combined centrally. &
Study-level results are aggregated. &
Researchers run code/queries inside the TRE; outputs are checked before release. \\
\midrule
\textbf{Participation} &
All data contributors must share data with the central site beforehand. &
Multiple data holders, dynamic participation possible (devices can join/leave). &
Centrally managed workers/nodes with data partitions. &
Fixed set of completed/published studies. &
Approved users/projects with strict governance and access control. \\
\midrule
\textbf{Data assumptions} &
No inherent assumption; depends on chosen analysis method. &
Must handle non-IID data and uneven sample sizes. &
Often assumes roughly IID, evenly partitioned data. &
Models between-study heterogeneity (fixed/random effects). &
No inherent assumption; depends on chosen analysis method. \\
\midrule
\textbf{What moves across parties} &
Individual data sent to the central site. &
Model updates (gradients/weights), possibly in shares (SMPC), encrypted or differentially private. &
Data blocks and intermediate results. &
Study-level summary statistics. &
Code/queries go in; vetted results come out. \\
\midrule
\textbf{Privacy posture} &
Highest data exposure (requires trust in central data custodian). &
Designed to avoid individual data sharing; can support privacy- enhancing techniques. &
Not privacy-focused (single trust domain). &
Only summary results shared. &
Via technical or organisational controls. \\
\midrule
\textbf{Output artefact} &
Single trained model or analysis result from pooled data. &
Global or personalised model held by each participant. &
Finished job outputs. &
Summary results with uncertainty estimates. &
Analysis outputs are released after disclosure control. \\
\midrule
\textbf{Typical examples} &
Central data warehouse, pooled data in a multi-centre study. &
Cross-hospital FL; edge device FL. &
Spark, Hadoop, Ray, HPC clusters. &
Cochrane-style meta-analyses. &
UK Biobank RAP, Federated EGA. \\
\midrule
\textbf{Legal responsibilities} & The central data controller has the responsibility for legal compliance and security. &
Data controllers retain responsibility for legal compliance and security; data processors have contractual responsibilities linked to that. &
Depends on data origin: same as centralised learning for single-centre studies, or as federated learning when data comes from multiple centres.
&
Data controllers retain responsibility; data processors must ensure original data use agreements permit meta-analysis. &
The operator is responsible for TRE security and governance. Data controllers retain legal responsibility for sharing the data. \\
\midrule
\textbf{Legal basis in addition to data subjects' consent } & Only data subjects' informed consent is needed. &
Data sharing agreements for pseudo-anonymised data. &
Data sharing agreement for individual data in case data from multiple centres are aggregated.
&
No personal data involved if the data are sufficiently aggregated (anonymised); otherwise, same as federated learning. &
Access agreements between TREs and data controllers: permitted uses, audit, security protocols. Data processors' agreements with TRE. \\

\bottomrule
\end{tabular}
}
\end{table}

\subsection{Topologies}
\label{sec:topologies}

\begin{figure}[t]
  \centering
  \includegraphics[width=0.930\linewidth]{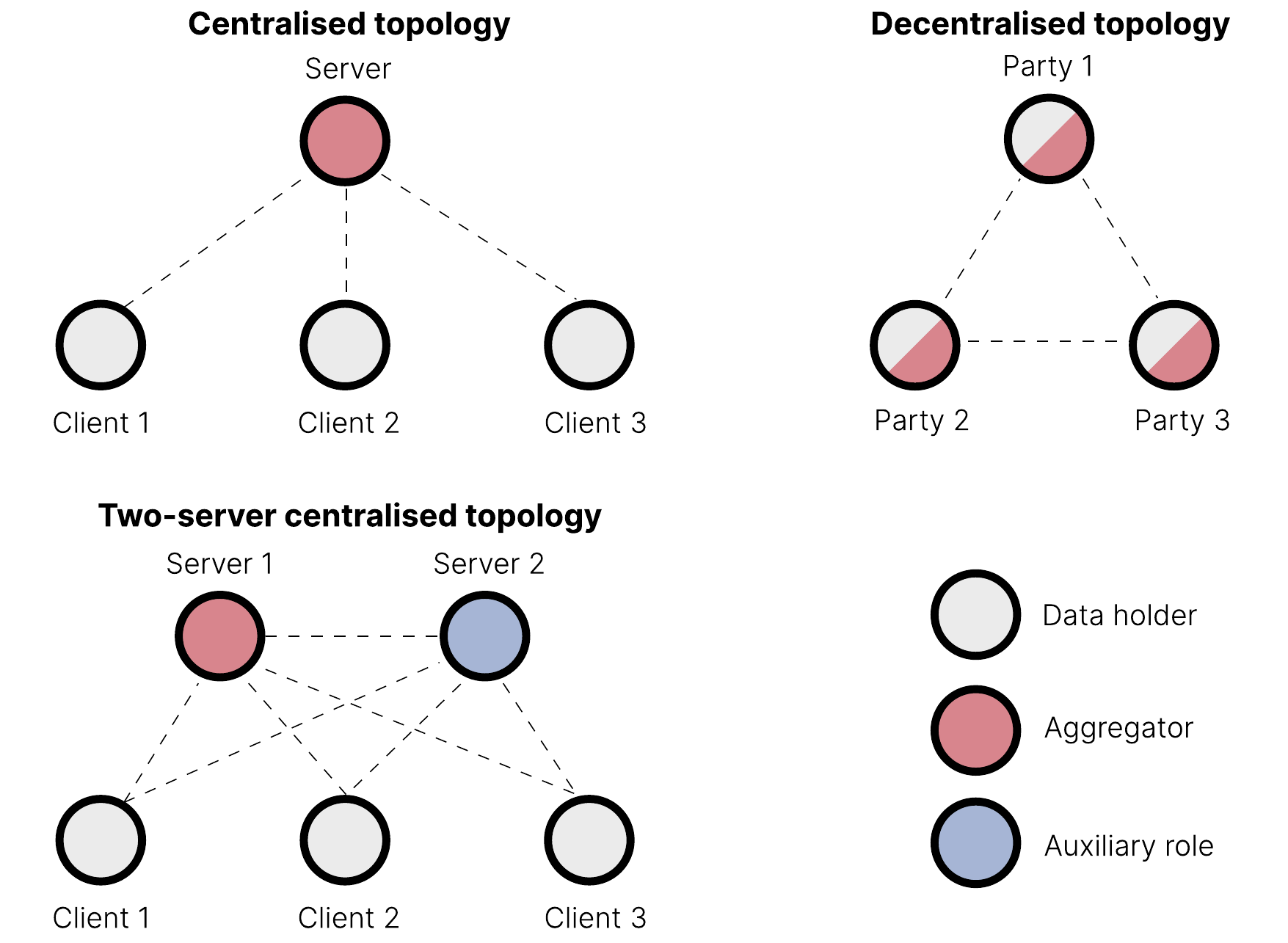}
  \caption{Different FL topologies. In centralised topologies, the data holders
    are typically referred to as \emph{clients}, reflecting their interaction
    with a central server. In decentralised topologies, where no central entity
    exists, the participants are often called \emph{parties}.}
  \label{fig:topology}
\end{figure}

The \emph{topology} of the FL consortium is determined by the number of
participating parties and their defined interactions. Some examples are
illustrated in Figure~\ref{fig:topology}. The most common is the
\emph{centralised} topology, where multiple data-holding parties (the
\emph{clients}) collaboratively train a shared machine learning model through a
central server (the \emph{aggregator}) that iteratively collects model updates
from each client, updates the global model, and redistributes it back to the
clients. Typically, clients do not communicate directly; they only communicate
with the central server. In contrast, a \emph{decentralised} topology
\citep{beltran} lacks a dedicated aggregation server. All consortium parties can
potentially serve as model trainers and aggregators, interacting through
peer-to-peer communication. Hybrid configurations include, for instance, using
two servers: one server handles aggregation of noisy local models, while the
other performs auxiliary tasks, such as noise aggregation \citep{hyfed}. Clients
can communicate with the servers, and servers can communicate with each other,
but clients cannot communicate with each other.

We will focus on the standard centralised topology and its two-server variant
here because, to our knowledge, no bioinformatics applications use decentralised
topologies.

\subsection{Hardware and software}
\label{sec:infra}

Hardware, software and models should be chosen with knowledge of the data and
inputs from domain and machine learning specialists to design an effective
machine learning pipeline \citep{pragmatic}.

In terms of infrastructure, FL requires computational resources for each client
and server. The optimal hardware configuration depends on the models to be
trained; at a minimum, each client must be able to produce model updates from
local data, and each server must be able to aggregate those updates and manage
the consortium. Connection bandwidth is not necessarily critical: to date,
client-server communications contain only a few megabytes of data, reaching
150MB only for large computer vision models, and can be made more compact
through compression and model quantisation \citep{camajori}. On the other hand,
latency may be a bottleneck if it limits the hardware utilisation.

As for software, several dedicated FL frameworks, many of which are
comparatively analysed in \citep{riedel}, provide structured tools and
environments for developing, deploying, and managing federated machine learning
models. While some frameworks, such as Tensorflow Federated \citep[TFF;][]{tff},
specialise in particular models, others support a broader range of approaches.
Notable open-source examples include PySyft \citep{pysyft} and Flower
\citep{flower}. Both are supported by active communities and integrate with
PyTorch to train complex models. PySyft is a multi-language library focusing on
advanced privacy-preserving techniques, including differential privacy and
homomorphic encryption. Flower is an FL framework: its modular design and ease
of customisation make it particularly useful for large-scale and multi-omics
studies involving heterogeneous devices and clients. We will provide examples
using these frameworks in Section~\ref{sec:bioinfofl} before discussing
frameworks explicitly designed for bioinformatics in
Section~\ref{sec:specialised}.

Other frameworks target healthcare and biomedical applications, but not
bioinformatics specifically. For instance, OpenFL \citep{foley} is designed to
facilitate FL on sensitive EHRs and medical imaging data; it supports different
data partitioning schemes (Section~\ref{sec:partitioning}) but struggles with
heterogeneous cross-device FL (Section~\ref{sec:allocations}). NVIDIA Clara,
which was used in \citet{covid19}, has similar limitations.

\subsection{Usage scenarios: cross-device and cross-silo}
\label{sec:allocations}

FL applications take different forms in different domains. Many small,
low-powered clients, such as wearable medical devices from the Internet of
Things, may produce the data needed to train the federated machine learning
model. Such \emph{cross-device} communications are often unreliable: passing
lightweight model updates instead of individual data largely addresses
connectivity issues and privacy risks.

FL may also involve a small number of parties, each possessing large amounts of
sensitive data \citep{huang22}, stored within their "data silos". This setting,
often called the emph{cross-silo} scenario, is common in healthcare and
bioinformatics. Here, the main priority is to minimise the privacy risks
associated with data sharing and comply with regulations. Additionally,
minimising large data transfers is also computationally advantageous when
modelling large volumes of information, such as whole-genome sequences.

These two scenarios differ in how they handle model updates. In the cross-silo
scenario, all (few) data holders in the consortium must participate in each
update. In contrast, we can rely on a subset of (the many) data holders in the
cross-device scenario because each holds a smaller share of the overall data.
This article focuses on the cross-silo scenario, as nearly all bioinformatics
applications fall within this framework.

\subsection{Data partitioning and heterogeneity}
\label{sec:partitioning}

\begin{figure}[t]
  \centering
  \includegraphics[width=0.930\linewidth]{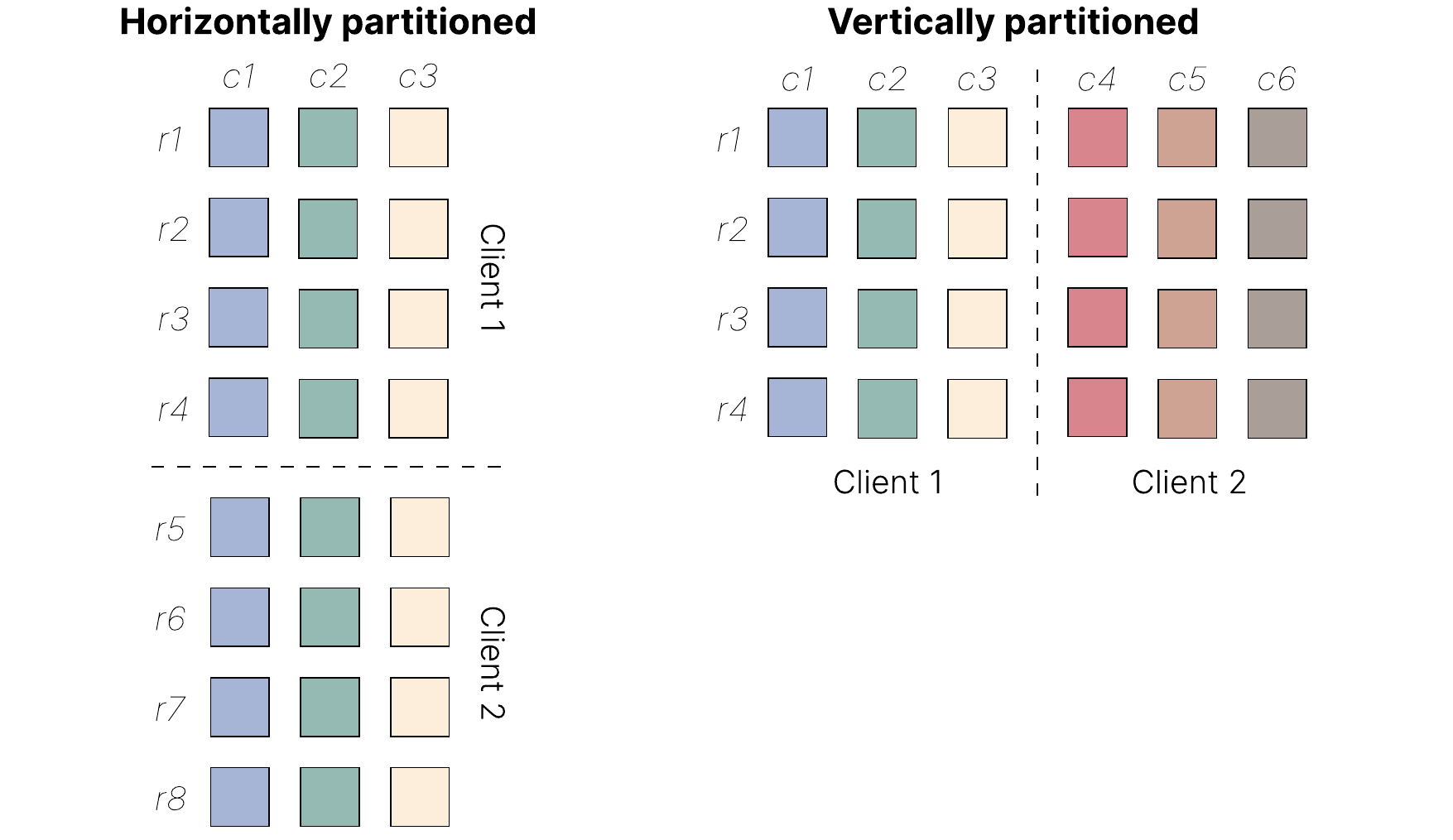}
  \caption{Horizontal and vertical data partitioning in FL. In horizontal FL
    (left), clients hold data sets with the same features (\emph{c1--c3}) but
    different subsets of samples (\emph{r1--r8}). In vertical FL (right),
    clients hold data sets with different features (\emph{c1--c6}) but the same
    set of samples (\emph{r1--r4}).}
  \label{fig:partitioning}
\end{figure}

Data may be partitioned along two axes: each party may record the same features
for different samples or features describing the same samples
(Figure~\ref{fig:partitioning}). In the first scenario, known as
\emph{horizontal} FL, different parties may each possess genomic sequencing data
from different individuals. In contrast, in \emph{vertical} FL, one party may
hold data from one omic type (say, genomic data), while another may have data
from a different phenotype or omic type (say, proteomic data) for the same
individuals. Horizontal FL is by far the most prevalent approach in
bioinformatics.

Significant variations in sample size and feature distributions between data
holders often exist. This heterogeneity allows FL to better capture the
variability of the underlying population, resulting in transferable models that
generalise well \citep{sheller}. Clearly, if data holders collect observations
from distinct populations, any federated model trained from them must be
correctly specified to capture population structure and avoid bias in inference
and prediction. If the populations are known, we can train targeted
population-specific models alongside the global one \citep{tan22}. Otherwise, we
can use clustering to identify them from the available data \citep{clusteredfl}.
Accounting for variations in measurements, definitions and distributions to
harmonise data across parties is also fundamental but is much more challenging
because access to data is restricted, even more so than in meta-analysis
\citep{camajori}.

\subsection{Security and privacy}
\label{sec:security}

FL reduces some privacy and security risks by design by passing model updates
between parties instead of centralising data in a single location. However, it
does not eliminate them completely.

In terms of privacy, deep learning models are the most problematic in machine
learning because of their ability to memorise training data. They leak
individual observations during training \citep[through model updates;
][]{geiping}, after training \cite[through their parameters;][]{haim} and during
inference \citep[membership attacks;][]{shokri,hu22}. More broadly, individual
reidentification is an issue for genetic data \citep{homer} and all the models
learned from them. For instance, \citep{cai} has demonstrated that it is
possible to identify an individual from the linear model learned in an
association study from just 25 genes. However, such works make unrealistic
assumptions on the level of access to the models and the data \citep{wainakh}:
even basic infrastructure security measures and the distributed nature of the
data will make such identification difficult under the best circumstances. The
privacy-enhancing techniques discussed in Section~\ref{sec:privacy} can make
such efforts wholly impractical.

As for security, we must consider different \emph{threat models}, understanding
what information requires protection, their vulnerabilities, and how to mitigate
or respond to threats. Internal and external threats to the consortium should be
treated equally with \emph{security in depth} design and implementation
decisions that consider parties untrusted. Security threats, such as membership
attacks and model inversion attacks \citep{inversion}, can originate equally
from parties and external adversaries that seek to abuse the model inference
capabilities to extract information about the data. On the other hand,
adversarial attacks are more likely to originate from consortium parties that
seek to introduce carefully crafted data or model updates into the training
process to produce a global model with undesirable behaviour. Some examples are
data poisoning \citep{poisoning}, manipulation \citep{manipulation} and
Byzantine attacks \citep{byzantine}.

Encrypting communication channels, implementing strict authentication (to verify
each party's identity) and authorisation (to control which information and
resources each party has access to or shares) schemes, and keeping comprehensive
access logs for audit can secure any machine learning pipeline, including
federated ones. Similarly, using an experiment tracking platform makes it
possible to track data provenance, audit both the data and the training process
and ensure the reproducibility of results \citep{pragmatic}. These measures must
be complemented by federated models resistant to these threats at training and
inference time, as thoroughly discussed in \citet{yin21}.

\subsection{Privacy-enhancing techniques}
\label{sec:privacy}

Privacy-enhancing techniques improve the confidentiality of sensitive
information during training. We summarise the most relevant below, illustrating
them in Figure~\ref{fig:example}.

Homomorphic encryption \citep[HE;][]{gentry} is a cryptographic technique that
enables computations to be performed directly on encrypted data (ciphertexts)
without requiring decryption. The outcome of operations on ciphertexts matches
the result of performing the same operations on the corresponding non-encrypted
values (plaintexts) when decrypted. HE can be either \emph{fully homomorphic}
(FHE), which allows for arbitrary computations, or \emph{partially homomorphic}
(PHE), which supports only a specific subset of mathematical operations. For
instance, the Paillier PHE scheme \citep{paillier} only supports additive
operations on encrypted data. FHE requires considerable computational resources
for encryption and decryption. PHE is less flexible but computationally more
efficient, making it a common choice in practical applications.

Secure multiparty computation  \citep[SMPC;][]{zhao19} is a peer-to-peer
protocol allowing multiple parties to compute a function over their data
collaboratively, similarly to Figure~\ref{fig:topology} (centre). Each data
holder divides their data into random shares and distributes them among all
parties in the consortium, thus ensuring that no single party can access the
complete data set. The shares are then combined during the computation process,
often with the assistance of a server, to produce the correct result while
preserving data privacy. SMPC ensures high security with exact results and keeps
data private throughout the computation process. However, SMPC is
computationally intensive and requires peer-to-peer communication, leading to
high communication overhead. Its complexity also increases with the number of
participants, limiting scalability.

Another approach to securing FL is using an aggregator and a compensator server
in a centralised two-server topology \citep[Figure~\ref{fig:topology},
right;][]{hyfed}. Each client adds a noise pattern to their local data, sharing
the former with the compensator (which aggregates all noise patterns) and the
latter with the aggregator (which aggregates the noisy data and trains the
model). The aggregator then obtains the overall noise pattern from the
compensator and removes it from the aggregated noisy data, allowing for denoised
model training. This two-server approach is efficient: it requires neither
extensive computation in the clients nor peer-to-peer communication. However, it
makes infrastructure more complex and requires trust in both servers not to
collude to compromise the privacy of individual contributions.

Unlike the above methods, which are encryption-based methods ensuring data
confidentiality during transmission or storage, differential privacy, another
popular technique for data-protection in federated learning, is not an
encryption system but rather a technique that focuses on privacy by ensuring
that the output of data analysis does not leak sensitive information about the
underlying dataset. Differential privacy \citep[DP;][]{ficek} achieves this
through a mathematical framework designed to ensure analyses remain
statistically consistent, regardless of whether any specific individual's data
is included or excluded. This property guarantees that sensitive information
about individuals cannot be inferred from the results up to a preset "privacy
budget" worth of operations. DP is typically implemented by introducing noise
into the data \citep{dp-data,dp-data2},  weight clipping in the training process
\citep{dp-sgd,dp-train} or predictions \citep{dp-pred,dp-pred2} to obfuscate
individual contributions. The amount of noise must be carefully calibrated to
balance predictive accuracy and privacy within the analysis: too little noise
undermines privacy, and too much reduces performance. This effect is more
pronounced within specific subgroups underrepresented in the training set
\citep{bagdasaryan}.

\begin{figure}[t]

% Define a coloured panel environment
\begin{tcolorbox}[
    %colback=LightCyan!10,  % Light background color
    %colframe=Teal,         % Frame color
    %coltitle=Black,        % Title color
    %fonttitle=\bfseries,   % Bold title font
    boxrule=0.3mm,          % Border thickness
    arc=4mm,                % Rounded corners
    width=\textwidth,     % Full width
    left=2mm,               % Inner left margin
    right=2mm,              % Inner right margin
    top=2mm,                % Inner top margin
    bottom=2mm              % Inner bottom margin
]

\small

\textbf{Example: privacy-preserving sum in FL}

In this simple example, three clients, with values \(5\), \(10\), and \(15\),
respectively, aim to securely calculate their sum, which has a true value of
\(5  + 10 + 15 = 30\). We show how to compute this sum using three techniques
described in Section \ref{sec:privacy}.

\smallskip

\textbf{Homomorphic Encryption}
\begin{itemize}
  \item A trusted entity generates a public-private key pair and distributes the
    public key to the clients.
  \item Each client encrypts their value using the public key and an additive
    homomorphic encryption scheme: \(E(5)\), \(E(10)\), and \(E(15)\), where
    \(E(x)\) denotes the homomorphic encryption of \(x\).
  \item Clients send the encrypted values \(E(5)\), \(E(10)\), and \(E(15)\) to
    the server.
  \item The server performs homomorphic addition on the encrypted values: \(E(5)
    + E(10) + E(15) = E(30)\).
  \item The aggregated encrypted value \(E(30)\) is sent back to the trusted
    entity with access to the private key.
  \item Using the private key, the trusted entity decrypts \(E(30)\), obtaining
    \(30\).
\end{itemize}

\textbf{Secure Multiparty Computation}
\begin{itemize}
  \item Clients split their values into random shares as \(\{2; 1; 2\}\), \(\{3;
    3; 4\}\), and \(\{5; 5; 5\}\) respectively, and then send the first two
    shares each to one of the other two clients.
  \item Clients sum the received shares and their local share to obtain \(10\),
    \(9\), and \(11\) respectively, and then send the obtained values to the
    server.
  \item The server sums the received values, obtaining \(30\).
\end{itemize}

\textbf{Two-Server Approach}
\begin{itemize}
  \item Clients generate large random noise values, \(543\), \(2612\), and
    \(1633\), respectively.
  \item Clients add the noise to their respective data, obtaining \(548\),
    \(2622\), and \(1648\), and send these values to the aggregator server.
  \item Clients send their noise values to the auxiliary server.
  \item The auxiliary server calculates the total noise, \(4788\), and sends it
    to the aggregator server.
  \item The aggregator server computes the total of the noised contributions,
    \(4818\), and subtracts the total noise, \(4788\), obtaining \(30\).
\end{itemize}

\end{tcolorbox}

\caption{Example of privacy-preserving sum computation in FL using three
  different techniques. Note that although differential privacy is described in
  Section \ref{sec:privacy}, it is not included in this example, as it would not
  be suitable for such a calculation.}

\label{fig:example}

\end{figure}

\section{Federated learning in bioinformatics}
\label{sec:bioinfofl}

Most FL literature focuses on general algorithms and is motivated by
applications other than bioinformatics, such as digital twins for smart cities
\citep{ramu}, smart industry \citep{zhang} and open banking and finance
\citep{long}. Even the clinical literature mainly focuses on different types of
data and issues \citep{covid19,saskia}. Here, we highlight and discuss notable
examples of FL designed specifically for bioinformatics, summarised in
Table~\ref{tab:summary}. They are all in the early stages of development, so
their reliability, reproducibility, and scalability are open questions. However,
they hint at the potential of FL to perform better than meta-analysis and
single-client analyses on real-world data, comparing favourably to centralised
data analyses where data are pooled in a central location while addressing data
sharing and use concerns \citep{flimma,xu20}.

\subsection{Proteomics and differential gene expression}
\label{sec:diffexp}

Proteomics studies the complex protein dynamics that govern cellular processes
and their interplay with physiological and pathological states, such as cancer
\citep{maes}, to improve risk assessment, treatment selection and patient
monitoring. Differential expression analyses focus specifically on comparing
expression levels across different conditions, tissues, or cell types to
identify genes with statistically significant differences \citep{rodriguez}.

In addition to the issues discussed in Section~\ref{sec:foundations}, FL in
proteomics must overcome the challenge of integrating data from different
platforms \citep{rieke} while accounting for imbalanced samples and batch
effects. \citet{cai22} produced a federated implementation of DEqMS
\citep[FedProt;][]{zhu20} for variance estimation in mass spectrometry-based
data that successfully identifies top differentially-abundant proteins in two
real-world data sets using label-free quantification and tandem mass tags.

\citet{flimma} implemented a federated \emph{limma voom} pipeline \citep{voom}
on top of HyFed \citep{hyfed}, which uses the aggregator-compensator two-server
topology we described earlier. This approach was demonstrated on two extensive
RNA-seq data sets, proving robust to heterogeneity across clients and batch
effects. \citet{hannemann} trained a federated deep-learning model for cell type
classification using both Flower and TFF and different architectures, with
similar results.

\subsection{Genome-wide association studies}
\label{sec:gwas}

Genome-wide association studies (GWAS) aim to identify genomic variants
statistically associated with a qualitative (say, a case-control label) or
quantitative trait (say, body mass index). These studies mainly use regression
models, which can be largely trained using general-purpose federated regression
implementations with minor modifications to address scalability and correct for
population structure \citep[see, for instance, ][]{kolobkov}.

\citet{fedglmm} has developed the most complete adaptation of these models to
federated GWAS in the literature: it provides linear and logistic regressions
with fixed and random effects and accounts for population structure via a
genomic relatedness matrix. \citet{grmat} further provides a federated estimator
for the genomic relatedness matrix. Finally, \citet{fedgmmat} describes the
federated association tests for the genomic variants associated with this model.
All these steps incorporate HE to ensure privacy in the GWAS.

As an alternative, \citet{cho} built on REGENIE \citep{regenie} to avoid using a
genomic relatedness matrix and increase the scalability of GWAS while using MPC
and HE to secure the data. Despite the overhead introduced by the encryption,
this approach is efficient enough to work on a cohort of 401k individuals from
the UK Biobank and 90 million single-nucleotide polymorphisms (SNPs) in less
than 5 hours.

\begin{table*}[t]
\centering
\caption{Summary of key federated learning applications in bioinformatics}
\label{tab:summary}
\begin{tabular}{L{2cm} L{3.1cm} L{3.1cm} L{3.1cm} L{3.8cm}}
\toprule
\textbf{Field} & \textbf{Application} & \textbf{FL Methods} & \textbf{Data} & \textbf{References} \\
\midrule
\textbf{Proteomics and Differential Gene Expression} & Variance estimation, gene expression, cell type classification & FedDEqMS, FedProt, HyFed with limma voom, DL with Flower and TFF & Mass spectrometry, RNA-seq, 1-10k individuals and 10-100M biomarkers & \makecell[tl]{\citet{cai22}\\ \citet{flimma}\\ \citet{hannemann}\\ \citet{hyfed}} \\
\midrule
\textbf{GWAS} & Association testing, scalable regression & FedGLMM, federated GRM estimator, FedGMMAT, REGENIE with MPC/HE & SNPs, genotype and phenotype data, 2,5--275k individuals and 0.5--38M SNPs & \makecell[tl]{\citet{fedglmm}\\ \citet{grmat}\\ \citet{fedgmmat}\\ \citet{cho}} \\
\midrule
\textbf{Single-cell RNA-seq}& Cell type classification & scFed: ACTINN, SVM, XGBoost, GeneFormer & scRNA-seq from 2-55k cells and 1-2k genes & \makecell[tl]{\citet{wang24}\\ \citet{fedtree}} \\
\midrule
\textbf{Multi-omics} & Prognosis (cancer), diagnostics (Parkinson's) & Vertical FL, adaptive neural networks, benchmarking with Flower & Genomics, transcriptomics, proteomics, 100--1200 individuals and 100-700 features & \makecell[tl]{\citet{wang23}\\ \citet{danek}} \\
\midrule
\textbf{Medical Imaging} & Classification, segmentation, semi-supervised training & Federated labelling, harmonised feature learning & MRI, X-rays, histology images, 5--71k scans & \makecell[tl]{\citet{bdair}\\ \citet{yan}\\ \citet{jiang}\\ \citet{haggenmuller}\\ \citet{linardos}\\ \citet{yang}} \\
\midrule
\textbf{Specialised Tools} & FL software for bioinformatics workflows & sfkit, FeatureCloud & All the data above & \makecell[tl]{\citet{sfkit}\\ \citet{featurecloud}\\ \citet{berger}\\ \citet{froelicher21}} \\
\bottomrule
\end{tabular}
\end{table*}

\subsection{Single-cell RNA sequencing}
\label{sec:rnaseq}

Single-cell RNA sequencing (scRNA-seq) measures gene expression at the cellular
level, rather than aggregating it at the tissue level as in bulk RNA sequencing,
and identifies the distinct expression profiles of individual cell populations
within tissues \citep{hwang18,papalexi}.

\citet{wang24} developed scFed, a unified FL framework integrating four
algorithms for cell type classification from scRNA-seq data: the ACTINN neural
network \citep{ma20}, explicitly designed for this task; a linear support vector
machine; XGBoost based on \citet{fedtree}; and the GeneFormer transformer
\citep{theodoris}. They evaluated scFed on eight data sets evenly distributed
among 2--20 clients, suggesting that the federated approach has a predictive
accuracy comparable to that obtained by pooling the data and better than that in
individual clients. However, the overhead during training increases with the
number of clients, limiting the scalability to larger consortia. More recently,
\citet{bakhtiari2025fedscgen} introduced FedscGen, a federated implementation of
scGen \citep{lotfollahi2019scgen}, a variational autoencoder-based method for
batch effect correction. FedscGen employs secure SMPC for privacy-preserving
aggregation and achieves results that closely match those obtained under
centralised training.

\subsection{Multi-omics}
\label{sec:omics}

Proteomics, genomics, and transcriptomics capture different aspects of
biological processes. Integrating large data sets from different omics offers
deeper insights into their underlying mechanisms \citep{civelek}. Vertical FL
allows multiple parties to combine various features of the same patients into
multimodal omics data sets without exposing sensitive information \citep{liu24}.
For instance, \citet{wang23} trained a deep neural network with an adaptive
optimisation module for cancer prognosis evaluation from multi-omics data. The
neural network performs feature selection while the adaptive optimisation module
prevents overfitting, a common issue in small high-dimensional samples
\citep{rajput}. This method performs better than a single-omic analysis, but the
improvement in predictive accuracy is strongly model-dependent. Another example
is \citet{danek}, who built a diagnostic model for Parkinson's disease: they
provided a reproducible setup for evaluating several multi-omics models trained
on pre-processed, harmonised and artificially horizontally federated data using
Flower. Their study identifies a general but not substantial reduction in FL
performance compared to centrally trained models, which increases with the
number of clients and is variably affected by client heterogeneity.

\subsection{Medical imaging}
\label{sec:imaging}

Medical imaging studies the human body's interior to diagnose abnormalities in
its anatomy and physiology from digital images such as those obtained by
radiography, magnetic resonance and ultrasound devices \citep{suetens}. It is
the most common application of FL in the medical literature \citep{chowdhury22}.
As a result, protocols for image segmentation and diagnostic prediction are well
documented. Notable case studies target breast cancer \citep{breast}, melanomas
\citep{haggenmuller}, cardiovascular disease \citep{linardos}, COVID-19
\citep{yang,covid19}.

Machine learning applications that use medical imaging data typically face
challenges, including incomplete or inaccurate labelling and the normalisation
of images from different scanners and different protocols. \citet{bdair}
explored a federated labelling scheme in which clients produced ground-truth
labels for skin lesions in a privacy-preserving manner, improving classification
accuracy. \citet{yan} also proposed an efficient scheme to use data sets mainly
comprising unlabelled images, focusing on chest X-rays. Furthermore,
\citet{jiang} apply FL to learn a harmonised feature set from heterogeneous
medical images, improving both the classification and segmentation of histology
and MRI scans.

\subsection{Ready-to-use FL tools for bioinformatics}
\label{sec:specialised}

The need for user-friendly FL implementations of common bioinformatics workflows
has driven the creation of secure collaborative analysis tools
\citep{berger,froelicher21,wan22}. Two notable examples are sfkit and
FeatureCloud.

The sfkit framework \citep{sfkit} facilitates federated genomic analyses by
implementing GWAS, principal component analysis (PCA), genetic relatedness and a
modular architecture to complement them as needed. It provides a web interface
featuring a project bulletin board, chat functions, study parameter
configurations and results sharing. State-of-the-art cryptographic tools for
privacy preservation based on SMPC and HE ensure data protection
\citep{lattigo}.

FeatureCloud \citep{featurecloud} is an integrated solution that enables end
users without programming experience to build custom workflows. It provides
modules to run on the clients and servers in the consortium. Unlike sfkit,
FeatureCloud allows users to publish applications in its app store, including
regression models, random forests and neural networks. Developers must also
document how privacy guarantees are implemented in their apps.

\section{Practical insights on federation}
\label{sec:operations}

This section offers practical insights to help readers interested in building a
federated and secure analogue of an existing bioinformatics algorithm. We focus
on horizontal FL with the centralised topology from Figure~\ref{fig:topology}
(left). Consider $K$ different clients, each possessing a local data set $X^k$,
where $k = 1, \dots, K$. Each data set contains $n^k$ samples, denoted as
$x^k_{ij}$, where $i = 1, \dots, n^k$ represents the sample index, and $j = 1,
\dots, P$ represents the $P$ features for each sample. We denote a row (column)
of the matrix $X^k$ as $x^k_{i*}$ ($x^k_{*j}$). This describes a distributed
data set of $N = \sum_{k=1}^{K} n^k$ observations:
\begin{equation*}
  X = \left[  \begin{array}{c}
       X^1 \\
       X^2 \\
       ... \\
       X^K \\
  \end{array}\right] \,.
\end{equation*}

The following sections assume that an FL consortium has been established, the
necessary infrastructure is operational, and an appropriate FL framework has
been selected and installed. It is also assumed that a secure aggregation
protocol has been chosen, such as those described in Section \ref{sec:privacy}
and Figure \ref{fig:example}. The choice of a specific secure aggregation
protocol may depend on several factors, including technology and infrastructure
(e.g., the availability of a particular FL topology that drives the choice), as
well as privacy risks and scalability concerns, as discussed in Section
\ref{sec:privacy}. In the following sections, we provide a general overview of
sum-based mathematical operations built upon a secure aggregation protocol, as
well as operations involving federated averaging \citep[FedAvg;][]{fedavg}.

Coding examples using Flower \citep{flower} are available in our GitHub
repository (\url{https://github.com/IDSIA/FL-Bioinformatics}). We chose Flower
because it has a shallow learning curve for new FL users and provides a good
balance between simplicity and flexibility when implementing custom algorithms.
\citet{riedel} also identified Flower as a promising framework because it has a
large, active, and growing community of developers and scientists, as well as
extensive tutorials and documentation. In our examples, secure summation is
performed using the secure aggregation protocol \texttt{SecAgg+} \citep{bell}.
This protocol combines encryption with SMPC, using a multiparty approach in
which each client interacts with only a subset of the others. It is particularly
suitable for several FL contexts, as it is robust to client dropout and highly
scalable. In particular, a relevant aspect of the bioinformatics domain is that
it scales linearly with the size of the vectors to be aggregated \citep{li21}.

\subsection{Sum-based computations}

Let $a^k$ be real numbers stored by individual clients. We define the secure sum
of these numbers, performed through the selected secure aggregation protocol, as
$\Oplus_{k=1}^K a^k$. We can build on this simple, secure sum to construct a
wide range of operations. However, note that as the complexity of operations
increases, the amount of information revealed to the server may also increase.
Sum-based operations include:

\begin{itemize}
  \item The \emph{overall sample size} of the distributed data set as $N =
    \Oplus_{i=1}^K n^k$ from the local sample sizes $n^k$.

  \item The \emph{mean value of the $j$-th feature}, given $N$, as
    \begin{equation*}
      M_j = \frac{1}{N} \Oplus_{i=1}^K \left[ \sum_{i=1}^{n_k} x_{ij}^k \right].
    \end{equation*}
    Each client computes the inner sum on their local data, whereas the outer
    one is a secure sum aggregated across clients by the server.

  \item The \emph{variance of the $j$-th feature}, given $N$ and $M_j$, as
    \begin{equation*}
      V_j = \frac{1}{N-1} \Oplus_{k=1}^K \left[ \sum_{i=1}^{n_k} (x_{ij}^k - M_j)^2 \right],
    \end{equation*}
    which can be used to standardise the $j$-th feature as $(x_{*j}^k - M_j) /
    \sqrt{V_j}$.

  \item The \emph{Pearson correlation coefficient} of two features $j$ and $j'$,
    given $M_j$ and $M_{j'}$, as
    \begin{equation*}
      \rho_{j,j'}=\frac{\frac{1}{N-1}\Oplus_{k=1}^K \sum_{i=1}^{n_k} (x_{ij}^k - M_j)(x_{ij'}^k - M_{j'})}
           {\sqrt{V_j V_{j'}}} \,.
    \end{equation*}

  \item The \emph{matrix $X^T X$}, as $X^T X = \Oplus_{k=1}^K (X^k)^T X^k$,
    where $\Oplus$ is a secure element-wise sum. This matrix is equivalent to
    the covariance matrix for standardised data sets and is commonly used for
    PCA.

\end{itemize}

Beyond these general-purpose examples, many operations specific to
bioinformatics pipelines also rely on simple sums. These operations are often
straightforward generalisations or compositions of the examples introduced
above.

In differential gene expression studies, for instance, filtering out weakly
expressed genes is standard practice. Weakly expressed genes can be defined as
those whose expression values fall below a specified threshold $t$ in, for
instance, 70\% of the samples. Let $v^k$ be a vector belonging to client $k$,
where each vector component represents the number of samples in which the
expression level of the gene (e.g., counts) exceeds the threshold $t$. The
server can securely calculate $v = \frac{1}{N} \Oplus_{k=1}^K v^k$ and identify
weakly expressed genes as those whose corresponding components of $v$ are
smaller than 0.7.

A fundamental preliminary step in a GWAS is identifying the minor allele and its
frequency. Let $a^k$, $c^k$, $g^k$, and $t^k$  be vectors belonging to client
$k$, where each component corresponds to a specific SNP. The components of
$a^k$, $c^k$, $g^k$, $t^k$ represent the number of samples in which nucleotides
$A$, $C$, $G$, $T$ are observed, respectively. The server can securely compute
the aggregated allele counts across all clients as $a = \Oplus_{k=1}^K a^k$ and
similarly $c$, $g$, $t$ (where $t$ can also be computed by difference from $N$
and the other three vectors). For each SNP, the minor allele is determined by
comparing the corresponding components of $a$, $c$, $g$, $t$: the allele with
the smaller value is designated as the minor allele. This operation is crucial
because the minor allele within a single client's population may differ from the
minor allele when considering the whole distributed data set. Ensuring a
consistent definition of the minor allele across all clients is essential for
reliable downstream analyses.

\subsection{Federated averaging computations}

FedAvg is a widely used algorithm for training deep neural networks in FL. It
iteratively computes a weighted average of model parameters across clients, with
weights proportional to the local sample sizes $n^k$. Thus, it can be applied to
any parametric model, including linear models.

FedAvg proceeds as illustrated in Figure \ref{fig:intro}. The server first
broadcasts an initial global model with parameters $w_0$. At each step of the
algorithm, clients start with the global model $w_t$ and perform local updates
to produce updated local models $w_{t+1}^k$. The global model is updated after
each round of local training as the weighted sum of the local models:
\begin{equation*}
  w_{t+1} = \Oplus_{k=1}^{K} \frac{n^k}{N} w_{t+1}^k \,,
\end{equation*}
where we use the secure sum $\Oplus$ for aggregation (FedAvg is itself a
sum-based operation). After aggregation, the updated global model is distributed
back to the clients.

However, many bioinformatics pipelines rely on linear models rather than deep
learning models. One commonly used model is logistic regression, which is
applied in tasks such as gene expression analysis and GWAS. A federated
implementation of logistic regression can be achieved by starting with a
standard implementation and applying FedAvg, which aggregates the local models
after a specified number of iterations performed by the local logistic
regressions.

\section{Legal aspects of federated learning}
\label{sec:legal}

The legal frameworks used within FL consortia are rarely discussed in the
literature. \citet{ballhausen} describes both the technical and legal aspects of
a European pilot study implementing a federated statistical analysis by secure
multiparty computation. They established agreements between parties similar to
those between participants in a multi-centre clinical trial, as using SMPC and
exchanging model gradients was legally considered data pseudonymisation (rather
than anonymisation). FL was determined to require the same level of data
protection as regular data sharing, which is also the most conservative course
of action suggested in \citet{truong,lieftink}. All clients jointly controlled
the consortium and were responsible for determining the purpose and means of
processing, including obtaining approval from the respective Ethics Committees.
\citet{sun} similarly describes the server in their consortium as a trusted and
secure environment, supported by a legal joint controller agreement between the
data owners.

Following \citet{ballhausen}, establishing an FL consortium could be expected to
require all participating and involved parties to execute agreements that
regulate their interactions, the so-called DPA or DSA (data
protection/processing/sharing agreements). Doing so will establish the level of
trust between parties and their responsibilities towards each other, third
parties, and patients. Risk aversion suggests that it should include a
data-sharing clause to allow for the sharing of information, similar to a
centralised analysis, as described in Figure~\ref{fig:intro}. No party has
access to the data of other parties. Still, it is theoretically possible that,
in some cases, the model updates shared during FL could be deanonymised by
malicious internal or external attackers \citep{truong}. Parties may then be
reluctant to treat that information as non-personally identifiable without
formal mathematical proof of anonymisation and prefer to establish data
protection responsibilities with a data-sharing agreement. In the EU (GDPR), but
also other jurisdictions (national data protection laws), ``all the means
reasonably likely to be used should be considered to determine whether a natural
person is identifiable'' \citep{gdpr}. Securing infrastructure in depth using
best practices from information technology, defensive software engineering, and
data by secure computing and encryption can make malicious attacks impractical
with current technologies \citep[security by design and by
default;][]{volini,martin}. In addition, when FL involves models other than deep
neural networks, if the contributions of individual parties are well balanced
across the consortium and include a sufficiently large number of individuals,
the information exchanged may very well be the same summary information
routinely published as supplementary material to academic journal publications
\citep{jegou}. A recent systematic literature review of privacy attacks in FL
has also highlighted that many of them are only feasible under unrealistic
assumptions \citep{wainakh}. Therefore, reducing the amount of information
shared during FL and using secure computing must be considered to provide
increased protection against data leaks and misuse. Advertising such measures as
a key feature of the FL consortium will make partners and patients more
comfortable with contributing to federated studies \citep[see, for
instance,][]{ballhausen}. \citet{lieftink}, which investigated how FL aligns
with GDPR in public health, also acknowledges that FL mitigates many privacy
risks by enforcing purpose limitation, data use and information exchange
minimisation, integrity and confidentiality (at a cost, as discussed in Section
\ref{sec:partitioning}).

Furthermore, consortium parties must agree on how to assign intellectual
property (IP) rights. Bioinformatics research often has practical applications
in industry, which may involve patenting the results and apportioning any
financial gains arising from their use. Parties in the consortium jointly
control it and should share any gains from it \citep{bigdata}. FL consortia are
no different in this respect. From a technical standpoint, watermarking
techniques for tracking data provenance and plagiarism have been adapted to FL
\cite{tekgul} to identify data and model theft.

Additionally, the agreement establishing the FL consortium must outline its
relationships with third parties and their corresponding legal obligations.
Third parties that have access to the infrastructure may be required to sign a
data processing agreement to guarantee the safety and privacy of data. In many
countries in Europe, as well as in the US, patients have the right to withdraw
their consent to use their data at any time. This, in turn, may require
implementing procedures to remove individual data points from future federated
analyses.

We summarised these considerations in
Table~{\ref{tab:centralized_fl_dc_meta_tre}}, along with the key differences
from the alternatives we discussed in Section~\ref{sec:foundations}. FL provides
increased protection against data and model leaks, which should reduce the
perceived risk for parties and patients in contributing to the consortium.
However, out of an abundance of caution, establishing a consortium-wide
data-sharing agreement may help allocate and reduce party responsibilities in
the event of a privacy breach. The use of FL has a limited impact on other legal
aspects of collaborative analysis, such as IP handling and requirements for
third parties, because it is a technical solution that does not change the
fundamental legal rights and responsibilities of the parties involved in the
consortium. Table \ref{tab:legal} summarises the key legal and procedural steps
required to implement federated learning in biomedical research, as detailed
above.

\begin{table}[ht!]
\caption{Overview of legal, procedural, and technical actions in federated
  learning, with relevance to governance, data, and software.}
\label{tab:legal}
\centering
\tiny
\renewcommand{\arraystretch}{1.4}
\begin{tabular}{>{\raggedright\arraybackslash}p{2.cm}
>{\raggedright\arraybackslash}p{3.cm}
>{\raggedright\arraybackslash}p{2.5cm}
>{\raggedright\arraybackslash}p{2.5cm}
>{\raggedright\arraybackslash}p{5.cm}}
\hline
\textbf{Action} & \textbf{Documentation} & \textbf{Data controllers responsabilities} & \textbf{Data processors responsibilities} & \textbf{Notes} \\
\hline
\textbf{Ethical approval} &
Study protocol (including data sharing information). &
Obtain approval from local ethics committees. &
-- &
Making the use of secure computing transparent (thus limiting the ways data can be processed) should support ethics committees' trust in the approach. \\
\hline
\textbf{Consortium setup} &
Cooperation/ data protection/ processing/ sharing agreements &
Joint controllers, responsible for the purposes and means of processing and for data security and purpose adherence. &
Contractual responsibilities stemming from controllers' data security and purpose adherence responsibilities. &  Although parties cannot access each other's data, agreements often permit the sharing of private information to address potential leakages from shared model parameters. \\
\hline
\textbf{Data Protection Officer (DPO) advice} &
GDPR-compliant documentation, Data Protection Impact Assessment (DPIA), software documentation. &
Appoint DPO if needed under GDPR (Article 35) and obtain advice. &
- &
Consortium agreements should specify how DPO responsibilities are allocated; a single DPO may be designated at the consortium level or appointed from a subset of partners.
  \\
\hline
\textbf{Clinical data collection (retrospective or prospective)} &
Informed consent forms, data collection protocols. &
Collect consent from all patients. Guarantee the right of withdrawal. &
-- &
Data should generally be considered pseudo-anonymised.
Explaining secure computing methods to patients should foster trust and informed participation. \\
\hline
\textbf{Code deployment} &
Software license, deployment agreements. &
Grant local software deployment compliant with deployment agreements. &
Guarantee the software and platform's security and usage comply with the purposes and licences. &
The consortium should agree on where deployment happens (e.g., trusted
execution environment - TEE), what is deployed and how deployments are authorised. \\
\hline
\textbf{Intellectual Property (IP)} &
IP ownership agreements, licensing terms, publication policies. &
If the trained model is protected by IP rights, its ownership and usage are governed by the terms of the contract. &
Do not automatically participate in IP generated from the trained model. &
In FL, the trained global model is typically viewed as a joint IP artefact. There is ongoing work to establish IP allocation models, licensing templates, and tailored governance mechanisms. \\
\hline
\textbf{Model Governance} &
Model versioning logs, audit trails, validation reports, access control policies. &
Guarantee model trustworthiness &
Facilitate model governance and auditability&
While stringent requirements are requested only in case of model deployment, analytical or research-only contexts still require principled governance to ensure reproducibility, accountability, and ethical compliance.  \\
\hline
\end{tabular}
\end{table}

We now proceed to discuss how FL facilitates compliance with the key
requirements of GDPR \cite{gdpr} through its architecture. Data providers, which
have a complete control over and a more intimate knowledge of the data they
collected as well as a direct connection to data subjects, act in a "data
controller" role (Articles 4 and 24), taking "appropriate technical and
organisational measures" (Article 25) to ensure privacy and security, thus
minimising the risk of data breaches. Therefore, they can directly scrutinise
their use, notify data subjects about it to request consent (Article 9); allow
them to withdraw their data (Article 7); ensure lawful, fair and transparent
processing (Articles 12--15); and directly assess risks to data subjects and
minimise them through appropriate legal agreements. Consortium parties, which
include both data controllers (as clients) and data processors (as servers,
compute facilities, as defined in Article 4), are also required to use the
techniques described in Section~\ref{sec:security} to ensure data security and
privacy beyond what is provided by base FL (privacy by design, Articles 24, 25
and 32). For the same reasons, FL facilitates compliance with the EU AI Act
\cite{eu-ai}. Some of its requirements strengthen those in the GDPR, such as
data minimisation, localisation, transparency, auditability, security, and data
quality. Additionally, the EU AI Act requires efforts to mitigate bias, ensure
the robustness of models, implement human oversight, and assess high-risk
systems. Data controllers are in the best position to ensure these requirements
are met. Collectively, they can provide more representative samples that are
less prone to bias and fairness issues. Finally, the presence of multiple data
controllers in the consortium also implies that these requirements are verified
by several independent parties.

In contrast, the US AI Act is more flexible in its requirements, which are left
to sector-specific agencies to define and enforce. It focuses more on party
self-regulation and harm remediation rather than universal legal mandates and
prevention. As a result, cross-border EU-US consortia should rely on the EU-US
Data Protection Framework \citep{dpf} or put in additional safeguards as
required by the GDPR (Article 46). Even so, FL's privacy stance naturally fits
well with the practical implementation of the US AI Act.

\section{Challenges and future directions}

Federated learning (FL) has evolved rapidly over its relatively short lifetime,
becoming a widely adopted methodology across diverse domains. In a recent
article \citep{daly2025}, which includes several authors of the seminal FL paper
\citep{fedavg}, the progress of the field is reviewed, and key challenges for
its future development are outlined. The authors propose a refined definition of
FL centred on privacy principles and analyse how core concepts such as data
minimisation, anonymisation, transparency and control, verifiability, and
auditability have evolved and are expected to play a major role in the future.
They identify three primary challenges for the field: scaling FL to support
large and multimodal models, overcoming operational difficulties arising from
device heterogeneity and synchronisation constraints, and addressing the current
lack of verifiability in deployed systems. As a potential means to address this
latter aspect, the authors highlight trusted execution environments (TEEs)
\citep{tee2015} as a promising technology. A TEE is a secure area within a
processor that executes code and processes data in isolation from other
software, protecting it from tampering or unauthorised access even if the main
operating system is compromised.

Complementing these insights, other surveys examine additional aspects of the
anticipated future developments in federated learning (FL) research.
\citet{wen2023survey} call for more efficient encryption schemes and greater
overall efficiency in FL, including strategies to reduce communication costs
between clients and servers. They also emphasise the importance of novel
aggregation strategies that can better handle client heterogeneity. In the
context of multimodal FL, aggregation must often reconcile contributions when
clients supply only partially overlapping modalities, which represents an
additional complexity beyond standard heterogeneity. \citet{yurdem2024}
highlight the emerging paradigm of FL as a Service, in which federated learning
training is conducted through ready-to-use platforms such as FeatureCloud
\citep{featurecloud} and sfkit \citep{sfkit}, discussed in
Section~\ref{sec:specialised}. This approach allows institutions to participate
in collaborative model development with minimal software and deployment effort,
thereby simplifying implementation and facilitating cross-organisational
collaboration. Looking ahead, the emergence of federated foundation models is
expected to define the next phase of research in this field. Their development
will require progress across several key dimensions, including improving
efficiency through advanced aggregation methods and optimised computational and
communication frameworks; strengthening trustworthiness by increasing robustness
to attacks; and enhancing incentive mechanisms that reward clients according to
the quality of the models provided. Collectively, these advances are expected to
be essential for making large-scale federated models practical and scalable
while maintaining manageable communication costs \citep{ren25}.

The prospects of federated learning in bioinformatics depend on how the legal
and technical landscape will develop. Further legislation will progressively
regulate the use of machine learning and AI models, defining and restricting how
data can be shared and used. As its effects percolate through protocol and
product development, many aspects of federated learning will likely take a more
definite shape.

Firstly, the security and privacy risks are likely to become more clearly
defined. Jurisprudence will naturally shift from general guiding principles,
such as the EU AI Act, to practical compliance rule sets as products based on
federated learning enter the market. How to assess sensitive aspects of data
(re)use will also likely be standardised \cite{gilbert}, using the vast
collection of available data and model cards as a starting point
\cite{chai,tripod-ai}. What threat models are relevant, what security measures
are appropriate at the infrastructure level, and what attacks are feasible will
then become clear, possibly confirming the irrelevance of many that have been
speculated in the literature \cite{wainakh}. The evolution of encryption and
differential privacy techniques may also allow different types of data to be
treated as anonymised, depending on their nature and the theoretical guarantees
of those techniques. Some data (say, single-cell transcriptomics) are
intrinsically more difficult to tie to a specific individual than others (say,
whole genome sequences), and different privacy-enhancing techniques are more
effective than others at mitigating various types of risk.

Secondly, the evolution of data and models will require periodic reevaluation of
the trade-offs discussed in this paper. The ever-increasing volumes of data used
to train federated models will eventually force a cost-benefit analysis for
resource-intensive techniques like HE and MPC; lighter approaches may be the
only feasible choice at scale, provided regulations permit their use. Modelling
approaches will undergo a similar selection process, as they will be required to
evolve to handle new biological and clinical data types in larger quantities and
across multiple modalities. In this respect, we foresee that federated learning
in bioinformatics may diverge from the mainstream, which has largely
standardised on deep learning models \cite{xu20,chowdhury22}. Historically,
bioinformatics has produced distinct model classes for different types of data.
Federating them, optimising them, and assessing their privacy risks will require
significant research and engineering efforts before they are suitable for
practical applications.

\section{Conclusions}
\label{sec:conclusions}

Independent research efforts in several bioinformatics domains have shown
federated learning to be an effective tool to improve clinical discovery while
minimising data sharing \citep[][among others]{breast,pati22,melloddy}. FL
enables access to larger and more diverse data pools, resulting in faster and
more robust exploration and interpretation of results. At the same time, it
provides enhanced data privacy and can seamlessly incorporate advanced,
encrypted and secure computation techniques. Despite the increased computational
requirements and reduced ability to explore and troubleshoot issues with the
data \citep{lieftink}, the benefits of FL may outweigh these additional costs.

Therefore, federated learning can potentially mitigate the risks associated with
national regulations if implemented in a manner that is secure by design and by
default. Its use may also make patients and institutions more confident in
participating in clinical studies by reducing privacy and data misuse risks.
However, the reliable use of federated learning and its effective translation
into clinical practice require a concerted effort by machine learning, clinical,
and information technology specialists. All their skills are necessary to
accurately evaluate the associated risks and expand its practical applications
in bioinformatics beyond the early-stage applications reviewed in this paper.

\section*{Conflict of interest statement}

The authors declare that the research was conducted without any commercial or
financial relationships that could potentially create a conflict of interest.

\section*{Author contributions}

DM: Conceptualization, Funding acquisition, Investigation, Methodology, Writing
  - original draft, Writing - review \& editing.

MS: Conceptualization, Investigation, Methodology, Writing - original draft,
Writing - review \& editing.

FG: Methodology, Writing - original draft, Writing - review \& editing.

HS: Writing - original draft, Writing - review \& editing.

AML: Funding acquisition, Project administration, Writing - original draft,
Writing - review \& editing.

IH: Writing - original draft, Writing - review \& editing.

MF: Conceptualization, Funding acquisition, Project administration, Writing -
original draft, Writing - review \& editing.

JvS: Writing - original draft, Writing - review \& editing.

SWvdL: Funding acquisition, Project administration, Writing - original draft,
Writing - review \& editing.

\section*{Funding}

This work was supported by the European Union Horizon 2020 programme
[101136962]; UK Research and Innovation (UKRI) under the UK Government's Horizon
Europe funding guarantee [10098097, 10104323] and the Swiss State Secretariat
for Education, Research and Innovation (SERI).

\bibliographystyle{Frontiers-Vancouver}

% \bibliography{references}

\end{document}